\documentclass[11pt]{revtex4-2}

\usepackage{hyperref}
\usepackage{braket}
\usepackage{amsmath,amsfonts,amssymb}
\hypersetup{dvips,dvipdfm,colorlinks=true,urlcolor=magenta,filecolor=magenta,linktoc=page,citecolor=red,linkcolor=blue,bookmarks=true}
\usepackage{graphicx,epstopdf}

\begin{document}
	\title{{A description of classical and quantum cosmology for a single scalar field torsion gravity}}
	\author{ Dipankar Laya$^1$\footnote {dipankarlaya@gmail.com}Roshni Bhaumik$^1$\footnote {roshnibhaumik1995@gmail.com} Sourav Dutta$^2$\footnote {sduttaju@gmail.com} Subenoy Chakraborty$^1$\footnote {schakraborty.math@gmail.com}}
	\affiliation{$^1$Department of Mathematics, Jadavpur University, Kolkata-700032, West Bengal, India\\$^2$Department of Mathematics, Dr. Meghnad Saha College, Itahar, Uttar Dinajpur-733128, West Bengal, India.}

	\begin{abstract}
		In the background of homogeneous and isotropic flat FLRW space-time, both classical and quantum cosmology has been studied for teleparallel dark energy (DE) model. Using Noether symmetry analysis, not only the symmetry vector but also the coupling function in the Lagrangian and the potential of the scalar field has been determined. Also symmetry analysis identifies a cyclic variable in the Lagrangian along the symmetry vector and as a result the Lagrangian simplifies to a great extend so that classical solution is obtained. Subsequently, in quantum cosmology Wheeler-DeWitt(WD) equation has been constructed and the quantum version of the conserved momenta corresponding to Noether symmetry identifies the periodic part of the wave function of the universe and as a result the Wheeler-DeWitt equation becomes solvable. Finally, quantum description shows finite non-zero probability at the classical big-bang singularity.  
	\end{abstract}
\maketitle

\section{Introduction}
Duality symmetry has a very important role in the development of conformal field theory. To address various issues in the early universe this symmetry technique has been used in string theory by Veneziano \cite{r1} and it is commonly known today as string cosmology. A dilaton field in the form of a scalar field $\phi(x^k)$ is coupled to gravity in string cosmology with action integral in the form \cite{r2, r3, r4, r5}:
\begin{equation}
S_{dil}=\int d^Dx\sqrt{-g}e^{-2\phi}\left[R-4g^{\mu\nu}\phi_{,\mu}\phi_{,\nu}+\Lambda\right]\label{1}
\end{equation}
with $\Lambda$ the usual cosmological constant. It is to be noted that the above action has some similarities with the BD-theory \cite{r6} and more specifically, one may get back to the Brans-Dicke theory \cite{r7} with constant BD parameter by the change of variable $\phi(x^k)\rightarrow-\frac{1}{2}\ln\psi(x^k)$. In the context of cosmology, string cosmology \cite{r19} may describe the inflationary scenario in a natural way without imposing any fine-tune potential. From perturbative view point, galactic magnetic fields corresponds to electro-magnetic perturbations while smallness of matter perturbation agrees with the homogeneity of the Universe \cite{r8}.

In the above action for the $D$--dimensional spatially flat and homogeneous background space-time, the Lagrangian depends on the scale factor as well as on the scalar field and interestingly there is a duality transformation\cite{r1}. :
\begin{equation}
a(t)\rightarrow a^{-1}(t)~,~\phi(t)\rightarrow\phi(t)-(D-1)\ln{a}\label{2}
\end{equation}
which leaves the above action integral to be invariant. Subsequently, the above duality transformation has been extended to anisotropic and inhomogeneous space-time models known as Gasperini-Veneziano duality property \cite{r2, r3, r4} and is specifically O(d,d) symmetry. Specifically, this duality symmetry is a discrete transformation and there exist an isometry in the background space-time. Subsequently, due to this discrete transformation one gets
\begin{equation}
H\rightarrow-H~,~t\rightarrow-t\mbox{~and~} \dot{H}(t)\rightarrow\dot{H}(-t),\nonumber
\end{equation} 
for FLRW model and hence one obtains the string-driven pre-big-bang cosmology. Further, the above scale factor duality transformation corresponds to a local transformation which leaves the action integral to be invariant (a variational symmetry\cite{r7}). Thus the above action integral (\ref{1}) can also describe the pre--big bang scenario.

In the present work, the above scale factor duality transformation has been used for teleparallel dark energy model. The introduction of modified gravity theory as well as the exotic matter is very much relevant in the context of cosmological observations for the last two decays or more. Further, teleparallel gravity has been attained special attention for the years due to a systematic geometric description to explain the cosmological observations. In this modified gravity theory, one gets the GR equivalence by considering the curvatureless Weitzenb$\ddot{o}$ck connection instead of the torsionless Levi-Civita connection. As a consequence the four linearly independent vierbeins are treated as dynamical fields and gravitational fields are defined by the Weitzenb$\ddot{o}$ck tensor and its scalar $T$. There is an analogus scalar-tensor theory with introduction of a scalar field in the teleparallel action integral, having interaction with the scalar $T$ for the Weitzenb$\ddot{o}$ck tensor. This modified gravity theory is also termed as scalar-torsion theory \cite{r9, r10, r11, r12}. The evolution of cosmological dynamics in teleparallel dark energy model has been investigated in \cite{r13, r14, r15} and with an analysis from observational view point in \cite{r16, r17}. The aim of the present work is to consider a teleparallel DE model having a discrete transformation from the point of view of Noether symmetry analysis. The plan of the work is as follows: 
Overview of teleparallel DE model is presented in Section-II and classical and quantum aspects of Noether symmetry is presented in Section-III. Section-IV deals with classical cosmology of teleparallel DE model, physical metric and symmetry analysis of this model is described in Section-V. Section-VI presents quantum cosmology and Noether symmetry analysis. The paper ends with a brief summary.

~~~~~~~~~~~~~~~~~~~~~~~~~~~~~~~~~~~~~~~~~~~~~~~~~~~~~~~~~~~~~~~~~~~~~~~~~~~~~~~~~~~~~~~~~~~~~~~~~~~~~~~~~~~~~~~~~~~~~~~~~~~~~~~~~~~~~~~~~~~~~~~~~~~~~~~~~~~~~~~~~~~~~~~~~~~~~~~~~~~~~~~~~~~~~~~~~~~~~~~~~~~~~~~~~~~~~~~~~~~~~~~~~~~~~~~~~~~~~~~~~~~~~~~~~~~~~~~~~~~~~~~~~~~~~~~~~~~~~~~~~~~~~~~~

 \section{An overview of Teleparallel gravity model}
	 In the teleparallel equivalent of general relativity (or teleparallel gravity in brief) instead of torsion, curvature is assumed to vanish and the underlying space-time is known as weitzenb$\ddot{o}$ck space-time. Though there is fundamental difference, yet the two theories are found to yield equivalent descriptions of the gravitational interaction i.e., both curvature and torsion provide an equivalent description of the gravitational interaction.
	In teleparallel gravity, the orthogonal tetrad components (i.e., vierbein field) $e_A (x^{\mu})$ form an orthonormal basis for the tangent space at each point $(x^{\mu})$ of the manifold i.e.,
		\begin{equation}
			e_Ae_B=\eta_{AB}=\mbox{diag}(+1, -1, -1, -1)\label{2.1}
		\end{equation} 
		are considered as  dynamical variables. So in a co-ordinate basis, the tetrad components can be expressed as
		\begin{equation}
			e_A=e_A^{\mu} \partial_{\mu}\label{2.2}
		\end{equation}
		where $e_A^{\mu}$ are the components of $e_A$, with $\mu=0, 1, 2, 3$ and $A=0, 1, 2, 3$. According to convention, the capital letters refer to the tangent space and co-ordiantes on the manifold are labeled by Greek indices. Thus considering the dual vierbein, one may write the metric tensor as a function of coordinates as 
		\begin{equation}
			g_{\mu \nu}(x)=\eta_{AB}e_{\mu}^A(x)e_{\nu}^B(x) \label{2.3}
		\end{equation}
		where $e^A(x^l)=h^A_{\mu}(x^l) dx^{\mu}$ is the dual basis with $e^A(e_B)=\delta_{B}^{A}$. { Here the original 16 degrees of freedom of the tetrad are constrained by the above 10 equations in equation(\ref{2.3}) and one has 6 new independent degrees of freedom. As the metric tensor has 10 degrees of freedom, so there are 4 new degrees of freedom to completely describe the theory. These are termed as the spin connection coefficients and are given by
		$${w_{\mu}}^{AB}={e_{\nu}}^A~{\Gamma^\nu}_{\sigma \mu}~e^{\sigma B}+{e_{\nu}}^A~\partial_{\mu}~e^{\nu B}$$
	From physical point of view, in canonical formulation of general relativity, a spin connection is defined on spatial slices and can be considered as the guage field generated by the local rotations.\\}
	In teleparallel gravity, the fundamental geometric object is the torsion tensor which is described by the antisymmetric part of the affine connection as
	\begin{equation}
		{T^{\sigma}_{\mu\gamma}=\Gamma^{\sigma}_{[\mu\gamma]}}\label{a1}
	\end{equation}
    Since the metric tensor is covariantly constant (i.e., ${\nabla_c{g_{ab}=0)},}$ {a generalized connection can be decomposed into a symmetric and an antisymmetric part as}
    \begin{equation}
    	{\Gamma_{\beta\gamma}^{\alpha}=\bar{\Gamma}_{\beta\gamma}^{\alpha}+k_{\beta\gamma}^{\alpha}}\label{a2}
    	\end{equation} 
    where the symmetric part $\bar{\Gamma}_{\beta\gamma}^{\alpha}$ is the usual christoffel symbols while the antisymmetric part $k_{\beta\gamma}^{\alpha}$ is known as contortion tensor with $k_{\alpha\beta\gamma}=k_{[\alpha\beta]\gamma}$. The interrelation between this contortion tensor and the above torsion tensor is 
    \begin{equation}
    	k_{\alpha\beta\gamma}=T_{\alpha\beta\gamma}+2T_{(\beta\gamma)\alpha}\label{a3}
    \end{equation}
	So torsion tensor can be consider as a connecting tool between the intrinsic angular momentum (spin) of the matter and the geometry of the space-time.\\
	{If the above antisymmetric connection for the teleparallel gravity is replaced by the Weitzenb$\ddot{\mbox{o}}$ck connection
	$\bar{\Gamma}_{\mu \nu}^{\sigma}=e^{\sigma}_{a}~\partial_{\mu}~e^{a}_{\nu}$ then the spin connection ${w_{\mu}}^{AB}$ vanishes identically. As a result} one may define torsion vector form the above torsion tensor due to its antisymmetric nature as
	\begin{equation}
		T_{\alpha}={T^{\beta}}_{\alpha\beta}=-{T^{\beta}}_{\beta\alpha}\label{a4}
	\end{equation}
    Hence one may expressed the contracted contortion tensor as 
    \begin{equation}
    	k^{\beta}_{\alpha\beta}=2T_{\alpha}=-{k_{\alpha\beta}}^{\beta}\label{a5}
    \end{equation} 
    Further, the torsion scalar $T$ can be define as 
    \begin{equation}
    	T={T^{\sigma}}_{\alpha\beta}S_{\sigma}^{\alpha\beta}\label{a6}
    \end{equation}
    where $S_{\sigma}^{\alpha\beta}=\frac{1}{2}\left({k^{\alpha\beta}}_{\sigma}+{\delta^{\alpha}}_{\sigma}{T^{\mu\beta}}_{\mu}-{\delta^{\beta}}_{\sigma}{T^{\mu\beta}}_{\mu}\right)$ be the super potential.
    
    In the present work, one considers the homogeneous and isotropic flat FLRW space-time having line-element 
    \begin{equation}
    	ds^2=-dt^2+a^2(t)\left[dr^2+r^2d\Omega_2^2\right]\label{a7}
    \end{equation} 
    where $a(t)$ is the scale factor and $d\Omega_2^2=d\theta^2+\sin^2\theta{d\phi^2}$ is the metric on unit 2-sphere.
    
    So the vierbein field in the diagonal form has the expression 
    \begin{equation}
    	h_{\mu}^{A}=\mbox{diag}(1, a(t), a(t), a(t))\label{2.8}
    \end{equation}
    with $T=6H^2$.

Now the action integral in teleparallel dark energy model can be written as
\begin{equation}
	S=\frac{1}{16\pi G}\int d^4x\sqrt{-g}\left[F(\phi)\left(T+\frac{\omega}{2}\phi_{,\mu}\phi^{,\mu} +V(\phi)\right)\right]\label{6}
\end{equation}
where $F(\phi)$ is a coupling function, $\omega$, a non-zero parameter similar to the Brans-Dicke parameter and $V(\phi)$ is the potential for the scalar field. Thus for the homogeneous and isotropic FLRW model the Lagrangian has the explicit form \cite{r5} 
\begin{equation}
	L(a,\dot{a},\phi,\dot{\phi})=F(\phi)\left[\frac{1}{N}\left(6a\dot{a}^2-\frac{\omega}{2}a^3\dot{\phi}^2\right)+Na^3V(\phi)\right]\label{7}
\end{equation} 
having field equations
\begin{eqnarray}
	F(\phi)\left(6H^2-\frac{\omega}{2N^2}\dot{\phi}^2-V(\phi)\right)&=&0\label{8}\\
	\left(\frac{2}{N}\dot{H}+3H^2\right)+\frac{1}{2}\left(\frac{\omega}{2N^2}\dot{\phi}^2-V(\phi)+2(\ln{F(\phi)})_{,\phi}H\frac{\dot{\phi}}{N}\right)&=&0\label{9}\\
	\omega\left(\frac{\ddot{\phi}}{N^2}+\frac{3}{N}H\dot{\phi}-\frac{\dot{N}}{N^3}\dot{\phi}+\right)+\left(\frac{\omega}{2N^2}\dot{\phi}^2+V(\phi)\right)+V_{,\phi}(\phi)&=&0\label{10}
\end{eqnarray}
where $N$ is the lapse function.

The basic difference of the present work with \cite{n1} is that multiscalarfield cosmology is considered in ref.\cite{n1} while present work deals with a self interacting scalar field (chosen as DE) in the formulation of teleparallel gravity. The previous work in ref.\cite{n1} can be reduced to the present work if $\beta=0$ and $\hat{V}=0$. The above field equations are highly coupled and non-linear in form. So it is very hard to find an analytic solution by solving the above set of differential equations. However, symmetry analysis will be used in the following sections not only to determine the cosmological solution but also to have a description of quantum cosmology.

~~~~~~~~~~~~~~~~~~~~~~~~~~~~~~~~~~~~~~~~~~~~~~~~~~~~~~~~~~~~~~~~~~~~~~~~~~~~~~~~~~~~~~~~~~~~~~~~~~~~~~~~~~~~~~~~~~~~~~~~~~~~~~~~~~~~~~~~~~~~~~~~~~~~~~~~~~~~~~~~~~~~~~~~~~~~~~~~~~~~~~~~~~~~~~~~~~~~~~~~~~~~~~~~~~~~~~~~~~~~~~~~~~~~~~~~~~~~~~~~~~~~~~~~~~~~

\section{Classical and Quantum Description using Noether Symmetry Analysis:a Brief Review}
	According to Noether's first theorem, if the Lagrangian of a physical system remains invariant with respect to the Lie derivative along an appropriate vector field \cite{r18} then there exist some conserved quantities associated with the physical system. The Euler-Lagrange equations take the form 
	\begin{equation}
			\partial_{j}\left(\frac{\partial L}{\partial\partial_{j}q^{\alpha}}\right)=\frac{\partial L}{\partial q^{\alpha}}\label{n1}
	\end{equation}
for a point-like canonical Lagrangian $L[q^{\alpha}(x^i),\dot{q}^{\alpha}(x^i)]$.

After contracting the equation (\ref{n1}) with some unknown function $\eta^{\alpha}(q^{\beta})$, one can get
	\begin{equation}
	\eta^{\alpha}\bigg[\partial_{j}\left(\frac{\partial L}{\partial\partial_{j}q^{\alpha}}\right)-\frac{\partial L}{\partial q^{\alpha}}\bigg]=0\label{n2}
\end{equation}
i.e,
\begin{equation}
	\eta^{\alpha}\frac{\partial L}{\partial q^{\alpha}}+(\partial_{j}\eta^{\alpha})\left(\frac{\partial L}{\partial\partial_{j}q^{\alpha}}\right)=\partial_{j}\left(\eta^{\alpha}\frac{\partial L}{\partial\partial_{j}q^{\alpha}}\right)\label{n3}
\end{equation}
Thus, the Lie derivative of the Lagrangian takes the form as
\begin{equation}
	\mathcal{L}_{\overrightarrow{X}}L=\eta^{\alpha}\frac{\partial L}{\partial q^{\alpha}}+(\partial_{j}\eta^{\alpha})\frac{\partial L}{\partial\left(\partial_{j}q^{\alpha}\right)}=\partial_{j}\left(\eta^{\alpha}\frac{\partial L}{\partial\partial_{j}q^{\alpha}}\right)\label{n4}
\end{equation}
This $\overrightarrow{X}$ is known as the infinitesimal generator of the Noether symmetry and it is defined by
\begin{equation}
	\overrightarrow{X}=\eta^{\alpha}\frac{\partial}{\partial q^{\alpha}}+\left(\partial_{j}\eta^{\alpha}\right)\frac{\partial}{\partial\left(\partial_{j}q^{\alpha}\right)}\label{n5}
\end{equation}
According to Noether's theorem, $\mathcal{L}_{\overrightarrow{X}}L=0$ which indicates the physical system to be invariant along the symmetry vector $\overrightarrow{X}$. It is to be noted that the Lagrangian and the symmetry vector are defined on the tangent space of the configuration $TQ\left\{q^\alpha,\dot{q}^\alpha\right\}$. Moreover, Noether symmetry analysis has a significant role to identify the conserved quantities of a physical system. From equation (\ref{n4}), one can say that there is a constant of motion of the system associated to this symmetry criteria. This conserved quantity is known as Noether current or conserved current which is defined as
\begin{equation}
	Q^i=\eta^{\alpha}\frac{\partial L}{\partial\left(\partial_{i}q^{\alpha}\right)}\label{n6}
\end{equation}
Also, this Noether current satisfies the condition 
\begin{eqnarray}
	\partial_iQ^i=0\label{n7}
\end{eqnarray}
The energy function which is defined by
\begin{equation}
	E=\dot{q}^{\alpha}\frac{\partial L}{\partial\dot{q}^{\alpha}}-L\label{n8}
\end{equation}
is a constant of motion if the Lagrangian of the system does not depend on time explicitly. Imposing these symmetry constraints to any physical system, the evolution equations of the physical system become solvable or simpler. 

In the context of quantum cosmology, Hamiltonian formulation is very much helpful and one can rewritten the Noether symmetry condition as
\begin{equation}
	\mathcal{L}_{\overrightarrow{Y}}H=0\label{n9}
\end{equation}
where ${\overrightarrow{Y}}=\dot{q}\dfrac{\partial}{\partial q}+\ddot{q}\dfrac{\partial}{\partial\dot{q}}$.

Due to Noether symmetry, the canonically conjugate momenta which is conserved in nature can be written as
	\begin{equation}
	p_{l}=\frac{\partial L}{\partial q^l}={\Sigma}_{l},~~~~l=1,2,...,m\label{n10}.
	\end{equation}
Here $m$ is the number of symmetries. Also the operator version of equation (\ref{n10}) can be written as
\begin{equation}
	-i\partial_{q^l}\ket{\psi}={\Sigma}_{l}\ket{\psi}\label{n11}.
\end{equation}
This equation (\ref{n11}) has oscillatory solution which is given by 
	\begin{equation}
	\ket{\psi}=\sum_{l=1}^{m}e^{i\sum_{l}q^l}\ket{\phi(q^k)}  , k<n.\label{n12}
\end{equation}
Here, $n$ is the dimension of the minisuperspace and $k$ is the direction where symmetry does not exist. The oscillatory part of the wave function indicates the existence of Noether symmetry and the conjugate momenta should be conserved along the symmetry vector and vice versa. This conserved momenta  due to Noether symmetry is useful to solve the Wheeler--DeWitt equation. In fact, using the separation of variables method it is possible to determine the non-oscillatory part of the wave function of the Universe and one may examine whether the initial big--bang singularity may be avoided or not by quantum description.

~~~~~~~~~~~~~~~~~~~~~~~~~~~~~~~~~~~~~~~~~~~~~~~~~~~~~~~~~~~~~~~~~~~~~~~~~~~~~~~~~~~~~~~~~~~~~~~~~~~~~~~~~~~~~~~~~~~~~~~~~~~~~~~~~~~~~~~
\section{Classical Cosmology of teleparallel DE Model}
To solving the above field equations (\ref{8})-(\ref{10}) we use Noether symmetry analysis. For the point-like Lagrangian (\ref{7}) of the given cosmological model the 
  infinitesimal generator for the Noether symmetry takes the form \cite{r21, r22, r23, r24, r25, r26}  
\begin{equation}
	\overrightarrow{X}=\alpha \frac{\partial}{\partial{a}} 
	+\beta \frac{\partial}{\partial{\phi}}+ 
	\dot{\alpha}\frac{\partial}{\partial{\dot{a}}}+ 
	\dot{\beta}\frac{\partial}{\partial\dot{\phi}}\label{11} 
\end{equation}
where $\alpha=\alpha(a,\phi),~\beta=\beta(a,\phi)$.

Now by imposing Noether symmetry to this Lagrangian (\ref{7}), we will get a system of partial differential equations : 
\begin{eqnarray}
	F(\phi)\alpha+a\beta F'(\phi)+2aF(\phi)\frac{\partial\alpha}{\partial{a}}&=&0\label{12}\\
	3F(\phi)\alpha+a\beta F'(\phi)+2aF(\phi)\frac{\partial\beta}{\partial{\phi}}&=&0\label{13}\\
	12\frac{\partial\alpha}{\partial\phi}-\omega a^2\frac{\partial\beta}{\partial a}&=&0\label{14}\\
	V'(\phi)+V(\phi)\left(\frac{3\alpha}{a\beta}+\frac{F'(\phi)}{F(\phi)}\right)&=&0\label{15}
\end{eqnarray}
The above set of partial differential equations can be solved by using the method of separation of variables and as a result the symmetry can have explicit form. So $\alpha,~\beta$ take the form  
\begin{eqnarray}
	\alpha\equiv\alpha(a,\phi)&=&\alpha_1(a)\alpha_2(\phi)\nonumber\\
    \beta\equiv\beta(a,\phi)&=&\beta_1(a)\beta_2(\phi)\label{16}
	\end{eqnarray}
As a result the explicit solutions are as follow  
\begin{equation}
	\alpha=\alpha_0a,\beta=\beta_0,~F(\phi)=F_0e^{-\frac{3\alpha_0}{\beta_0}\phi},~V(\phi)=V_0\label{17}
\end{equation} 
with $\alpha_0,~\beta_0,~F_0,\mbox{~and}~V_0$ as arbitrary constants. Thus symmetry analysis not only determine the symmetry vector but also the potential function and the coupling function.

\begin{figure}
		\centering{\includegraphics[height=7cm,width=10cm]{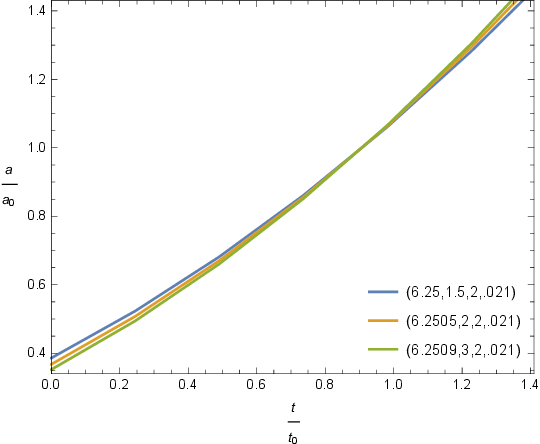}}
        \caption{Graphical representation of $\frac{a}{a_0}$ with respect to $\frac{t}{t_0}$ for different values of
        $(c_1,c_4,c_5,V_1)$ parameter spaces with $t_0=.011$.}\label{f1}
 \end{figure}

\begin{figure}
		\centering \includegraphics[height=7cm,width=10cm]{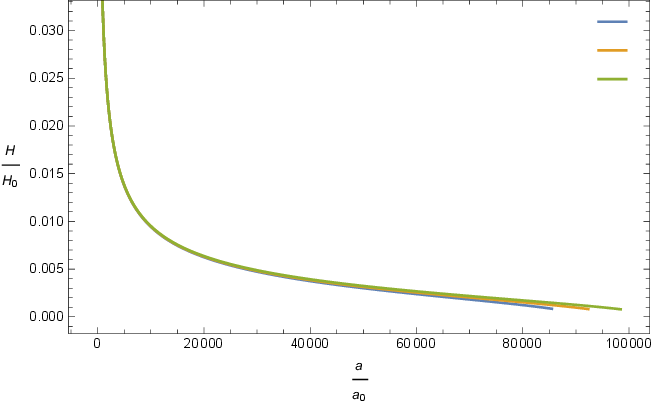}
        \caption{The graphical representation of  $\frac{H}{H_0}$ against  $\frac{a}{a_0}$ where $H_0=72.9798$.}\label{f2}
 \end{figure}

\begin{figure}
		\centering \includegraphics[height=7cm,width=10cm]{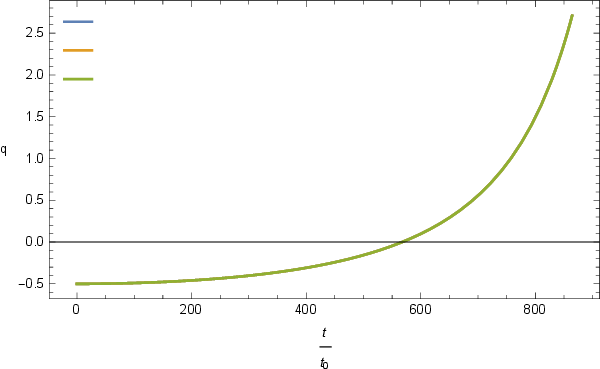}
	\caption{The graphical representation of deceleration parameter $q=-1-\frac{\dot{H}}{H^2}$ with respect to $\frac{t}{t_0}$.}\label{f3}
\end{figure}

Further, the symmetry vector helps us to identify the cyclic variable in the augmented space so that the Lagrangian as well as the field equations simplified to a great extend. So a transformation in the augmented space : $(a,\phi)\rightarrow(u,v)$ is chosen such that  
\begin{equation}
  i_{\overrightarrow{X}}du=1,~i_{\overrightarrow{X}}dv=0\label{18}  
\end{equation}
where the left hand side of the above equations indicate the inner product between the vector field $\overrightarrow{X}$ and the one form $du$ or $dv$. As a result the transformed symmetry vector field takes the form 
\begin{equation}
    \overrightarrow{X}_T=\frac{d}{du}\nonumber
\end{equation}
and the conserved current (i.e. conserved charge in the present case) can be written in compact form as
\begin{equation}
	Q=i_{\overrightarrow{X}}\theta_L\label{19}
\end{equation} 
with $\theta_L=\frac{\partial L}{\partial\dot{a}}da+\frac{\partial L}{\partial\dot{\phi}}\phi$, the Cartan one-form.
Thus from the transformation equation (\ref{18}) one has the explicit transformation in the augmented space as 
\begin{equation}
  u=\frac{\phi}{\beta_0},~v=\ln{a}-\frac{\alpha_0}{\beta_0}\phi.\label{20}
\end{equation} 
So the transformed Lagrangian in the new variables takes the form 
\begin{equation}
	L=F_0e^{3v}\left\{\left(6\alpha_0^2-\frac{\omega}{2}\beta_0^2\right)\dot{u}^2+6\dot{v}^2+12\alpha_0\dot{u}\dot{v}+V_{0}\right\}\label{21}
\end{equation} 
where $u$ is the cyclic coordinate. The corresponding Euler-Lagrange equations are 
\begin{eqnarray}
	(12\alpha_0^2-\omega\beta_0^2)\dot{u}+12\alpha_0\dot{v}&=&Ae^{-3v}\label{22}\\
    \mbox{and}~~
	2\ddot{v}+3\dot{v}^2+A_1e^{-6v}+V_1&=&0\label{23}
\end{eqnarray}

where $A_1,~V_1$ are connected to the relation $A_1=\frac{A^2}{8\omega\beta_0^2},~V_1=\frac{V_0(12\alpha_0^2-\omega\beta_0^2)}{4\omega\beta_0^2}$. Now we take $A=0$ to get a solution of above two equation. So after solving these two equations we get
\begin{eqnarray}
    u(t)&=&-\frac{8\alpha_0}{(12\alpha_0^2-\omega\beta_0^2)}\ln\left\{\cos\left(\frac{\sqrt{3V_1}}{2}(t-2c_1)\right)\right\}+c_3\label{24}\\
	v(t)&=&\frac{2}{3}\ln\left\{\cos\left(\frac{\sqrt{3V_1}}{2}(t-2c_1)\right)\right\}+c_2\label{25}
\end{eqnarray}
where $c_1,~c_2,~c_3$ are arbitrary constant. Now using previous relation (\ref{20})  we get the old coordinates
\begin{eqnarray}
	a(t)&=&c_4\left\{\cos\left(\frac{\sqrt{3V_1}}{2}(t-2c_1)\right)\right\}^{c_5}\nonumber\\
	\phi(t)&=&-\frac{8\alpha_0\beta_0}{(12\alpha_0^2-\omega\beta_0^2)}\ln\left\{\cos\left(\frac{\sqrt{3V_1}}{2}(t-2c_1)\right)\right\}+c_3\beta_0\label{26}
\end{eqnarray} 
where $c_4=e^{c_2+\alpha_0c_3},~c_5=\frac{2}{3}-\frac{8\alpha_0^2}{(12\alpha_0^2-\omega\beta_0^2)}$ i.e., $c_4$ is an arbitrary constant and $c_5$ is a constant. From the above classical solution the cosmological parameters namely the scale factor, Hubble parameter and the deceleration parameter have been shown graphically in figures (\ref{f1})-(\ref{f3}) for the various choices of the parameters involved. 


In FIG.\ref{f1} the dimensionless scale factor ($\frac{a}{a_0}$) has been plotted against the cosmic time ($\frac{t}{t_0}$) and the figure shows that the Universe is in an expanding phase throughout the evolution. The graphical representation of the dimensionless Hubble parameter ($\frac{H}{H_0}$) has been presented in FIG.\ref{f2}. The nature of the graph shows that the Hubble parameter is  positive throughout but decreasing in nature. So, the Universe is in expanding phase but the rate of expansion gradually decreases. The deceleration parameter has been plotted in FIG.\ref{f3}. The graph shows that the  present model describes the early inflationary era to the matter dominated era. So, this model is not able to describe the present accelerated expansion of the Universe. Further, it is to be noted that the nature of the scale factor and the Hubble parameter shown graphically (in FIG. (\ref{f1}) and(\ref{f2})) agree with the observational evidences since the very early era\cite{r26.1, r26.2}.

~~~~~~~~~~~~~~~~~~~~~~~~~~~~~~~~~~~~~~~~~~~~~~~~~~~~~~~~~~~~~~~~~~~~~~~~~~~~~~~~~~~~~~~~~~~~~~~~~~~~~~~~~~~~~~~~~~~~~~~~~~~~~~~~~~~~~~~~~~~~~~~~~~~~~~~~~

\section{PHYSICAL METRIC AND SYMMETRY ANALYSIS}
The Noether point symmetries of Lagrange equations having first order Lagrangian are shown to be generated by the elements of the homothetic group of the kinetic metric. In general if the field equations do not have Noether point symmetries then they are not Noether integrable. However, it is possible to have extra  Noether point symmetries by the above homothetic algebra.\\

For the present physical problem described by the Lagrangian (\ref{7}) one may define the kinetic space as a 2D space having line element
\begin{equation}
	d{S_{2}^{(k)}}^2=6aF(\phi)da^2-\frac{\omega}{2}a^3F(\phi)d\phi^2\label{27}
\end{equation}
having effective potential $V_{eff}=a^3F(\phi)V(\phi).$

As the present Lagrangian (in equation (\ref{7})) is in  the form of a point Lagrangian : $L=T-V_{eff}$, where $T$ can be considered as the K.E. of a point particle so the Noether point symmetries are generated by the elements of the homothetic group of the kinetic metric 
\begin{equation}
	d{S_{2}^{(k)}}^2=a^3F(\phi)\left[\frac{6}{a^2}da^2-\frac{\omega}{2}d\phi^2\right]\label{28}
\end{equation}
Thus we have associated conformal (1+1) decomposable metric as
\begin{equation}
	d{S_{2}^{(c)}}^2=\frac{6}{a^2}da^2-\frac{\omega}{2}d\phi^2=6du^2-\frac{\omega}{2}d\phi^2,~u=\ln{a}\label{29}
\end{equation}
Further, the above kinetic line-element has the usual gradient homothetic vector (HV) $H_V=\frac{2}{3}a\partial_a$ with $\psi H_{\psi}=1,$ but it does not generate a Noether point symmetry of the present Lagrangian. Moreover, the $2D$ metric in $(u,\phi)$-plane is a space of constant curvature having 2 Killing vectors which generate the So(2) group. In addition, the above conformally flat metric (\ref{29}) has $4\left(\frac{n(n+1)}{2}+1 \mbox{~for~}n=2\right)$ dimensional homothetic Lie algebra as 
\begin{eqnarray}
	&(i)&~\overrightarrow{H}_V=u\frac{\overrightarrow{\partial}}{\partial u}~\mbox{with}~\psi.H_V=1~~(\mbox{gradient} HV)\nonumber\\
	&(ii)&~\overrightarrow{K}^{(1)}=\cosh\phi\partial_{u}-\frac{1}{u}\sinh\phi\partial_{\phi}~~(\mbox{gradient Killing vector field})\nonumber\\
	&(iii)&~\overrightarrow{K}^{(2)}=\sinh\phi\partial_{u}-\frac{1}{u}\cosh\phi\partial_{\phi}~~(\mbox{gradient Killing vector field})\nonumber\\
	&(iv)&~~\overrightarrow{K}^{(4)}=\frac{\overrightarrow{\partial}}{\partial\phi}~~(\mbox{non-gradient (rotational) Killing vector, generate the $So(2)$ algebra})\nonumber
\end{eqnarray}
The gradient Killing functions corresponding to $\overrightarrow{K}^{(1)}$ and $\overrightarrow{K}^{(2)}$ are 
\begin{equation}
	g_1=u, g_2=u.\nonumber
\end{equation}
The above analysis shows that the determination of Noether point symmetries reduce to a problem of differential geometry.

~~~~~~~~~~~~~~~~~~~~~~~~~~~~~~~~~~~~~~~~~~~~~~~~~~~~~~~~~~~~~~~~~~~~~~~~~~~~~~~~~~~~~~~~~~~~~~~~~~~~~~~~~~~~~~~~~~~~~~~~~~~~~~~~~~~~~~~~~~~~~~~~~~~~~~~~~~~~~~~~~~~~~~~~~~~~~~~~~~~~~~~~~~~~~~~~~~~

\section{QUANTUM COSMOLOGY AND NOETHER SYMMETRY ANALYSIS}
The canonically conjugate momenta (for the transformed Lagrangian) due to Noether symmetry can be written as  
\begin{eqnarray}
	p_u&=&\frac{\partial{L}}{\partial\dot{u}}=F_0e^{3v}\left\{(12\alpha_0^2-\omega\beta_0^2)\dot{u}+12\alpha_0\dot{v}\right\}\label{30}\\
	p_v&=&\frac{\partial{L}}{\partial\dot{v}}=F_0e^{3v}\left\{12\dot{v}+12\alpha_0\dot{u}\right\}\label{31}
\end{eqnarray}
%with Hamiltonian for this model 
%\begin{equation}
%	H=F_0e^{3v}\left\{6\dot{v}^2+\left(6\alpha_0^2-%\frac{\omega}%{2}\beta_0^2\right)\dot{u}^2+12\alpha_0\dot{u}\dot{v}-%V_0\right\}
%\end{equation}
%Using (35),(36) then we get the results
%\begin{eqnarray}
%	\dot{u}&=&\frac{e^{-3v}}{\omega{F_0}\beta_0^2}(p_v\alpha_0-%p_u)\label{38}\\
%	\dot{v}&=&\frac{e^{-3v}}{12\omega{F_0}\beta_0^2}\left\{12\alpha_0p_u-(12\alpha_0^2-\omega\beta_0^2)p_v\right\}\label{39}
%\end{eqnarray}
%Substituting (38),(39) in (37) then
Then the Hamiltonian of the system can be written as 
\begin{equation}
	H=A_2e^{-3v}p_u^2+A_3e^{-3v}p_v^2-A_4e^{-3v}p_up_v-V_2e^{3v}\label{32}
\end{equation}
where $A_2,~A_3,~A_4$ and $V_2$ are arbitrary constants.

It is to be noted that as '$u$' is cyclic coordinate so the corresponding momentum $p_u$ is conserved i.e.,
\begin{equation}
	F_0e^{3v}\left\{(12\alpha_0^2-\omega\beta_0^2)\dot{u}+12\alpha_0\dot{v}\right\}=\mbox{conserved}=\sigma\label{33}
\end{equation}
 In quantum cosmology, the wave function of the Universe is a solution of the Wheeler-DeWitt(WD) equation, a second order hyperbolic partial differential equation. In fact, the WD equation is the operator version of the Hamiltonian constraint i.e., 
$\hat{H}\psi(u,v)=0$. But there is a problem of operator ordering in course of conversion to the operator i.e, $p_u \rightarrow -i \frac{\partial}{\partial u}$ and $p_v \rightarrow -i \frac{\partial}{\partial v}$. So the explicit form of the WD equation can be written as 
$$\left[A_2e^{-3v}\frac{\partial^2}{\partial{u^2}}+A_3e^{-3l_1v}\frac{\partial}{\partial v}e^{-3l_2v}\frac{\partial}{\partial v}e^{-3l_3v}-A_4\frac{\partial}{\partial u}e^{-3m_1v}\frac{\partial}{\partial v}e^{-3m_2v}+V_2e^{3v}\right]\psi(u,v)=0$$
where the number triplet $(l_1, l_2, l_3)$ and the doublet $(m_1, m_2)$ are arbitrary except the restrictions $l_1+l_2+l_3=0$ and $m_1+m_2=0$. Note that though there are infinite possible choices for the above triplet and doublet, still the following are the preferred choices namely\\

 $(i)~ l_1=2, l_2=-1,l_3=0, m_1=2, m_2=-1$ (D'Alembert operator ordering)\\
 
 $(ii)~ l_1=l_3=0,l_2=1, m_1=0, m_2=1$ (Vilenkin ordering)\\
 
 $(iii) ~l_1=1, l_2=l_3=0, m_1=1, m_2=0$ (No ordering)\\
 
It is to be noted that the issue of factor ordering influences the behaviour of the wave function, still at the semi classical level factor ordering has no effect\cite{r27}. Now due to simplicit of the form of the WD equation we shall restrict to the above third choice i.e, no ordering and as a result the above WD equation simplifies to  
\begin{equation}
\left[A_2e^{-3v}\frac{\partial^2}{\partial{u^2}}+A_3e^{-3v}\frac{\partial^2}{\partial{v^2}}-A_4e^{-3v}\frac{\partial^2}{\partial{u}\partial{v}}+V_2e^{3v}\right]\psi(u,v)=0\label{40}
\end{equation} 
As the wave function of the Universe is the general solution of the above WD equation so it can be expressed as a superposition of the eigenfunction of the WD operator i.e., 
$$\Psi(u, v)=\int W(\sigma) \psi (u, v, \sigma) d\sigma$$
where $\psi$ is an eigen function of the WD operator and $W(\sigma)$ represents the weight function which by proper choice gives us the desire wave packet. However, in analogy with classical description it is desirable to construct a coherent wave packet with good asymptotic behaviour in the minisuperspace, peaked around the classical trajectory.\\

Moreover, it is nice to examine whether the wave function can predict the evolution of the dynamical variables. For a consistent quantum cosmological model the classical solution can be predicted at late time only from the quantum description but quantum behaviour at early epochs is distinct from classical solution and is singularity free.\\

The operator version of conserved momentum which is given by equation (\ref{33}) takes the form
\begin{equation}
	-i\frac{\partial{\psi(u,v)}}{\partial{u}}=\sigma\psi(u,v)\label{37}.
\end{equation}
Solution of equation (\ref{37}) is  oscillatory in nature and it  indicates the existence of Noether symmetry.

From (\ref{37}), we get 
\begin{equation}
	\psi_1(u)=\sigma_0e^{i\sigma{u}}\label{39}
\end{equation}
where $\sigma_0$ is an arbitrary constant.

Putting the expression of $\psi_1(u)$ in the WD equation (\ref{40}), one can get a second order differential equation as 
\begin{equation}
	A_3\frac{d^2\psi_2(v)}{dv^2}-iA_4\sigma\frac{d\psi_2(v)}{dv}+V_2e^{6v}\psi_2(v)-A_2\sigma^2\psi_2(v)=0\label{41}
\end{equation}
 The solution of the above equation (\ref{41}) gives the non-oscillatory part of the wave function which is given bellow: 
\begin{equation}
	\psi_2(v)=e^{iA_5v}\left\{\mu_1\Gamma(1-A_6)J_{(-A_6)}(A_7e^{3v})+\mu_2\Gamma(1+A_6)J_{(-A_6)}(A_7e^{3v})\right\}\label{42}.
\end{equation}
where $\mu_1,~\mu_2$, $A_5$, $A_6$, $A_7$ are arbitrary constants $\Gamma$ is the usual incomplete Gamma function.  Therefore the wave function of the Universe for this model can be written as  
\begin{equation}
	\psi(u,v)=\sigma_0e^{i(\sigma{u}+A_5v)}\left\{\mu_1\Gamma(1-A_6)J_{(-A_6)}(A_7e^{3v})+\mu_2\Gamma(1+A_6)J_{(-A_6)}(A_7e^{3v})\right\}\label{43}
\end{equation}
Also in  $(a,\phi)$ coordinate system, the above wave function takes the form:
\begin{equation}
	\psi(a,\phi)=\sigma_0e^{i\left(\sigma{\frac{\phi}{\beta_0}}+A_5\left(\ln{a}-\frac{\alpha_0}{\beta_0}\phi\right)\right)}\left\{\mu_1\Gamma(1-A_6)J_{(-A_6)}(A_7a^3e^{-\frac{3\alpha_0}{\beta_0}\phi})+\mu_2\Gamma(1+A_6)J_{(-A_6)}(A_7a^3e^{-\frac{3\alpha_0}{\beta_0}\phi})\right\}\label{44}
\end{equation}

\begin{figure}[h]
	\centering \includegraphics[height=10cm,width=9cm]{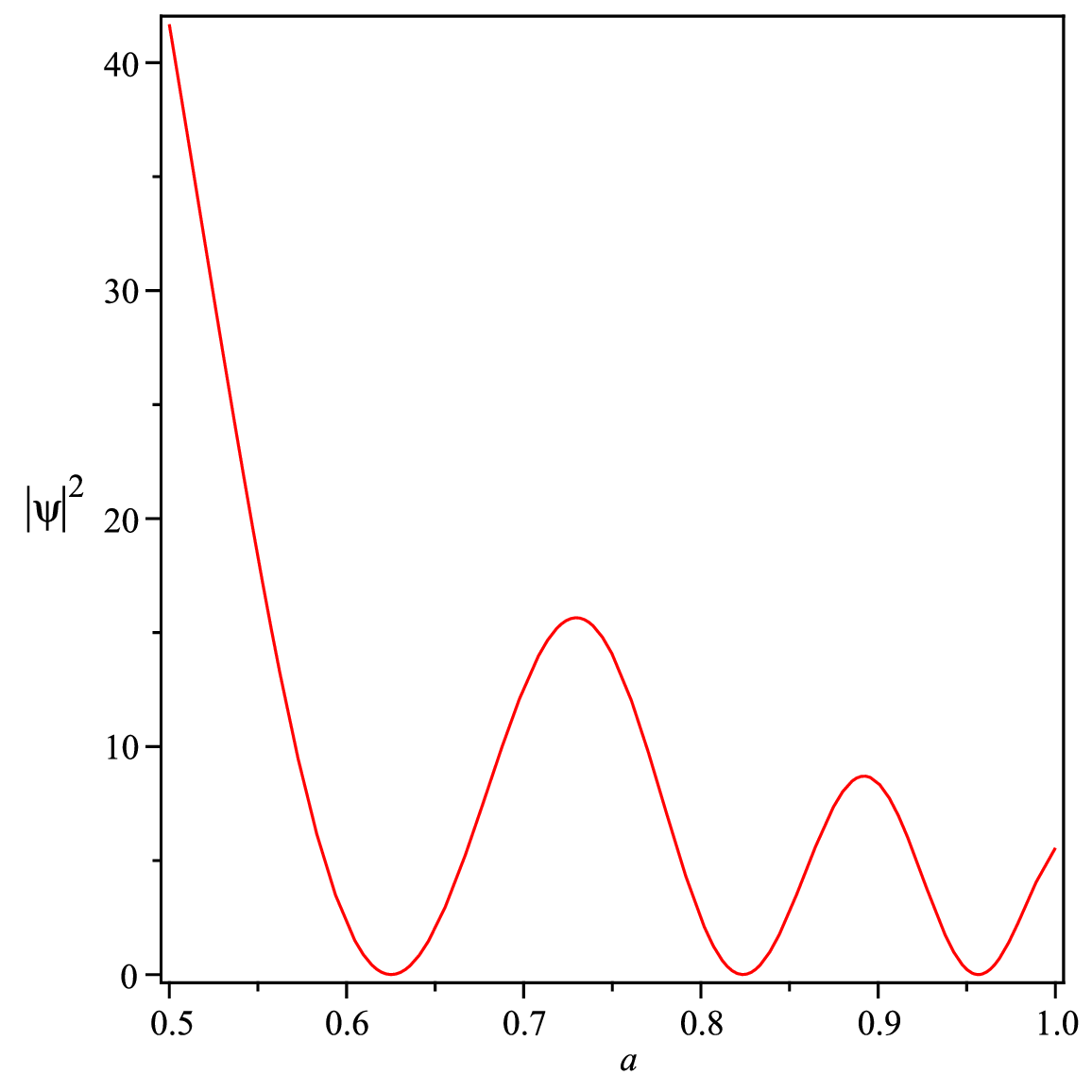}
	\caption{Wave function of the Universe}\label{f4}
\end{figure}

Here we have plotted $|\psi|^2$ with respect to the scale factor $a$ (FIG. \ref{f4}). From the graph it is clear that $|\psi|^2$ takes a non-zero finite value at zero volume. So this cosmological model may not overcome the big-bang singularity using quantum cosmology.

\section{Brief Summary}
The present work gives an illustration how symmetry analysis helps us to study both classical and quantum cosmology of a model Universe. Here Noether symmetry  analysis explores both classical and quantum cosmological description of teleparallel dark energy model. The Noether symmetry vector in the augmented space identifies a cyclic variable and as a result the field equations become solvable.
 The classical cosmological solution has been presented analytically in equation (\ref{26}) while the relevant cosmological parameters namely the scale factor, Hubble parameter and the deceleration parameter has been shown graphically in FIG (\ref{f1}-\ref{f3}). The graphical description shows that the Universe expands throughout the evolution while the rate of expansion gradually decreases as predicted by the recent observational evidences. However, the graphical representation of the deceleration parameter relates that the present cosmological model describes the early evolution of the Universe to the matter dominated era. In quantum cosmology WD equation (a second order hyperbolic partial differential equation) has been constructed  and the operator ordering issue has been discussed. The conserved momentum associated with the cyclic variable gives us a first order linear differential equation due to operator conversion and this first order differential equation gives us an oscillatory solution. Subsequently, it is possible to have solution of the WD equation with the help of the above periodic solution. The probability amplitude has been shown graphically in FIG.\ref{f4} and it is found that classical singularity can not be avoided by quantum description.\\
 
  For future work, we can extend the present work by choosing non--minimally coupled scalar field or by considering multi--scalar field model. Also it will be interesting to perform cosmological perturbations about the initial cosmic time.

\section*{Acknowledgments}
The author D.L. would like to thanks ``University Grants Commission, India" for helping the financial support. The NET-JRF Fellowship ID is 211610000075.


\frenchspacing
\begin{thebibliography}{58}
	\bibitem{r1}  G. Veneziano, {\it Phys. Lett. B} {\bf 265}, 287 (1991).

    
	\bibitem{r2} M. Gasperini and G. Veneziano, {\it Astropart. Phys.} {\bf 1}, 317 (1993).
	
	\bibitem{r3} T.H. Buscher, {\it Phys. Lett. B} {\bf 194}, 59 (1987).
	
\bibitem{r4} T.H. Buscher, {\it Phys. Lett. B} {\bf 201}, 466 (1988).
	
\bibitem{r5} A. Paliathanasis, {\it Eur. Phys. J. Plus} {\bf 136}, 674 (2021)
	
\bibitem{r6} C. Brans and R.H. Dicke, {\it Phys. Rev.} {\bf 124}, 925 (1961).

\bibitem{r7} A. Paliathanasis and S. Capozziello, {\it Mod. Phys. Lett. A} {\bf 31}, 1650183 (2016).

\bibitem{r19} G. Veneziano, Nucl. Phys. B (Proc. Suppl.) 55B, 134 (1997)


	
\bibitem{r8} L. McAllister and E. Silverstein, {\it Gen. Rel. Grav.} {\bf 40}, 565 (2008).

\bibitem{r9} M. Skugoreva, E.N. Saridakis and A. Toporensky, {\it Phys. Rev. D} {\bf 91}, 044023 (2015).
	
\bibitem{r10}  M. Hohmann, {\it Phys. Rev. D} {\bf 98}, 064002 (2018).

\bibitem{r11} M. Hohmann and C. Pfeifer, {\it Phys. Rev. D} {\bf 98}, 064003 (2018).

\bibitem{r12} M. Hohmann, {\it Phys. Rev. D} {\bf 98}, 064004 (2018).

\bibitem{r13} H. Wei, {\it Phys. Lett. B} {\bf 712}, 430 (2012).

\bibitem{r14} C. Hu, E.N. Saridakis and G. Leon, {\it JCAP} {\bf 07}, 005 (2012).

\bibitem{r15} G. Otalora, {\it JCAP} {\bf 07}, 044 (2013).

\bibitem{r16} C.-Q. Geng, C.-C Lei and E.N. Saridakis, {\it JCAP} {\bf 01}, 002 (2012).

\bibitem{r17} R. D’Agostino and O. Luongo, {\it Phys. Rev. D} {\bf 98}, 124013 (2018).

\bibitem{n1} K. Dialektopoulos, G. Leon, A. Paliathanasis {\it Eur. Phys. J. C} {\bf 83}, 218 (2023).

\bibitem{r18} S. Capozziello, A Stabile and A. Troisi, {\it Class. 
 Quant. Grav.} {\bf 24}, 2153 (2007) (ariv: gr-qc/0703067).

\bibitem{r21} A.Paliathanasis M.Tsamparlis. {\it Gen.Relt.Grav.}, {\bf 42}, 2957, (2010).

\bibitem{r22} A. Paliathanasis, M. Tsamparlis, S. Basilakos, 
 {\it Phys. Rev. D}, {\bf 91},123535 (2015).

\bibitem{r23} A. Paliathanasis, M. Tsamparlis, and S. Basilakos.  {\it Phys. Rev. D}, {\bf 90}, 103524 (2014).

\bibitem{r24}
A. Paliathanasis and G. Leon, {\it Class. Quantum Grav.} {\bf 38}, 075013 (2021).


\bibitem{r25} A. Paliathanasis and G. Leon, {\it Eur.Phys. J. Plus} {\bf 137}, 165 (2022).


\bibitem{r26} M. Tsamparlis and A. Paliathanasis, {\it J. Phys. A} {\bf 44}, 175202 (2011).

\bibitem{r26.1} A. G. Riess et.al., {\bf arXiv:2112.04510}.

\bibitem{r26.2} Planck Collaboration, {\bf arXiv:1807.06209} .

 \bibitem{r27} F. Tavakoli, B Vakili, {\it Gen. Rel. Grav.} {\bf 51}, 122 (2019).



\end{thebibliography}
\end{document}